\pdfoutput=1

\documentclass[a4paper]{aa} 
\usepackage[breaklinks, colorlinks, citecolor=blue]{hyperref}

\usepackage{graphicx}
\usepackage{epsfig}
\usepackage{array}

\usepackage{natbib}

\usepackage{amsmath, amssymb, stmaryrd}     
\usepackage[varg]{txfonts}  %

\newcommand{\noun}[1]{\textsc{#1}}

\begin{document}

\title{Cosmological simulations with disformally \\coupled symmetron fields}

\author{R. Hagala\thanks{E-mail: robert.hagala@astro.uio.no}
\and C. Llinares
\and D. F. Mota} 
\institute{Institute of Theoretical Astrophysics, University of Oslo, PO Box 1029 Blindern, 0315 Oslo, Norway}

\date{}

\abstract
{We investigate statistical properties of the distribution of matter at redshift zero in disformal gravity by using $N$-body simulations. The disformal model studied here consists of a conformally coupled symmetron field with an additional exponential disformal term.  }
{We conduct cosmological simulations to discover the impact of the new disformal terms in the matter power spectrum, halo mass function, and radial profile of the scalar field.}
{We calculated the disformal geodesic equation and the equation of motion for the scalar field. We then implemented these equations into the $N$-body code \noun{Isis}, which is a modified gravity version of the code \noun{Ramses}.}
{The presence of a conformal symmetron field increases both the power spectrum and mass function compared to standard gravity on small scales.  Our main finding is that the newly added disformal terms tend to counteract these effects and can make the evolution slightly closer to standard gravity.  We finally show that the disformal terms give rise to oscillations of the scalar field in the centre of the dark matter haloes.}
{}


\maketitle

\section{Introduction}

Since 1998, it has been known that the universe expands at an accelerating rate that is consistent with the existence of a cosmological constant $\Lambda$ \citep{1998AJ....116.1009R, 1999ApJ...517..565P}.  The standard model for cosmology, $\Lambda$CDM, gives an excellent fit to most modern precision observations of large scale structures and of the cosmic microwave background, but it does not explain what the source of $\Lambda$ is. Attempts to calculate the vacuum energy from particle physics yields answers that are several orders of magnitude off from the measured cosmological value of $\Lambda$. A cancellation of that many terms is very improbable and would require extreme fine tuning. This is known as the cosmological constant problem and is a severe problem in modern physics (see \citet{1989RvMP...61....1W} for an early introduction to this issue).

A viable solution to the cosmological constant problem is to assume that the particle physics vacuum energy is totally concealed on gravitational scales, while other mechanisms are responsible for the measured expansion. One way to search for such mechanisms consist of introducing a slight modification to standard general relativity (GR) in such a way that the equations for gravity will give rise to accelerated expansion on large scales. There are innumerable models for modified gravity (see \citet{2012PhR...513....1C} for a review).

To be considered viable, an important requirement that any modification of gravity must fulfil is that it should reduce to GR on solar system scales.  This is achieved through so-called screening mechanisms, described in detail by \citet{Joyce:2014kja}. Screening mechanisms are needed because conventional GR is tested and confirmed to very high precision on solar system scales, meaning any modifications to gravitational physics must give similar results within very tight constraints on these scales (see the review by \citet{2009aosp.conf..203R} for an overview over solar system tests and constraints).

In this paper we investigate a specific form for modified gravity by using $N$-body simulations, namely the disformal model of gravity. Disformal models were first introduced by \citet{1993PhRvD..48.3641B}, and have now been widely studied by applying them to inflation, dark energy, and dark matter \citep{dis1,dis2,dis3,dis4,dis5,dis6,dis7,dis8,dis9,dis11,dis14,dis12,dis13,dis10,2015JCAP...04..036V}. However, this is the first time disformal models have been studied on non-linear scales.

We consider a model where the scalar field has both conformal and disformal couplings to matter.  For the conformal part we use a symmetron potential \citep[][]{2005PhRvD..72d3535P, 2008PhRvD..77d3524O, 2010PhRvL.104w1301H}, while for the disformal part we use an exponential term.

Conformally invariant symmetron models have already been investigated in the non-linear regime \citep{2012ApJ...748...61D, 2014PhRvD..89h4023L,2014A&A...562A..78L}, as well as several other conformally invariant scalar tensor theories 
\citep{Boehmer:2007ut,Li:2008fa,baldi_coupled_quintessence_code, zhao_baojiu_forf_code,Li2,dgp_code_durham,nbody_chameleon, 2014PhRvL.112v1102H}.  
The aim of this paper is to investigate, for the first time, the effects of adding a disformal term to the symmetron field.  We will focus our analysis mainly on the statistical properties (i.e. the power spectrum and halo mass function) of the simulated matter distribution on non-linear scales.  The simulations are performed with a modified version of the $N$-body code \noun{Isis} \citep{2014A&A...562A..78L}, which is itself a modified gravity version of the code \noun{Ramses} \citep{{2002A&A...385..337T}}.

The paper is organised as follows: Section 2 describes the equation of motion of the scalar field and the associated geodesic equation; it also summarises how these are implemented into \noun{Isis} and tested.  Section 3 describes the cosmological simulations that are used for the analysis and Section 4 contains the results from the analysis.  We discuss the results and draw conclusions in Section 5.

\section{The equations and the code}

\subsection{The model}

The model is defined by the following scalar-tensor action
\begin{equation}
S=\intop\left[\sqrt{-g}\left(\frac{R}{16\pi G}-\frac{1}{2}\phi^{,\mu}\phi_{,\mu}-V\left(\phi\right)\right)+\sqrt{-\bar{g}}\bar{\mathcal{L}_{\mathrm{m}}}\right]\mathrm{d}^{4}x,\label{eq:action}
\end{equation}
where $g$ and $\bar{g}$ are the Einstein and Jordan frame metrics, $R$ is the Ricci scalar, and $\bar{\mathcal{L}}_{m}$ is the Lagrangian density of matter (computed using the Jordan frame metric $\bar{g}$ whenever applicable). The field potential $V\left(\phi\right)$ can have many different forms, but we choose the quartic symmetron potential with the three free parameters $\mu$, $\lambda$, and $V_{0}$, 
\begin{equation}
V\left(\phi\right)=-\frac{1}{2}\mu^{2}\phi^{2}+\frac{1}{4}\lambda\phi^{4}+V_{0}.
\end{equation}
In disformal gravity the Jordan frame metric $\bar{g}$ is related to the Einstein frame metric according to 
\begin{equation}
\bar{g}_{\mu\nu}=A\left(\phi\right)g_{\mu\nu}+B\left(\phi\right)\phi_{,\mu}\phi_{,\nu}.
\end{equation}
The specific forms of $A$ and $B$ that we will study in this paper are as follows:
\begin{align}
A\left(\phi\right) & = 1+\left(\frac{\phi}{M}\right)^{2},\\
B\left(\phi\right) & = B_{0}\exp\left(\beta\frac{\phi}{\phi_{0}}\right), 
\end{align}
where $B_{0}$ and $\beta$ are free parameters for the disformal coupling. The normalization constant $\phi_{0}$ is chosen to be the vacuum expectation value of the symmetron field $\phi_{0}\equiv\frac{\mu}{\sqrt{\lambda}}$. The mass scale $M$ is a free parameter, which decides the interaction strength of the conformal coupling.

This specific choice of $A$, $B$, and $V$ gives a symmetron model with an additional disformal term described by $B\left(\phi\right)$.  The model reduces to the symmetron model when fixing $B_{0}=0$.  Both the matter-coupled symmetron part $A$ and the disformal part $B$ can lead to two separate screening effects \citep{2010PhRvL.104w1301H, dis4}. We use natural units where $c=\hbar=1$.

\subsection{The equation of motion for the scalar field}

The equation of motion for the scalar field that results from varying the action defined in Eq. \eqref{eq:action} is:
\begin{multline}
\left(1+\gamma^{2}\rho\right)\ddot{\phi}+3H\dot{\phi}-\frac{1}{a^{2}}\nabla^{2}\phi= \\
\gamma^{2}\rho\left(\frac{A_{,\phi}\left(\phi\right)}{A\left(\phi\right)}\dot{\phi}^{2}-\frac{B_{,\phi}\left(\phi\right)}{2B\left(\phi\right)}\dot{\phi}^{2}-\frac{A_{,\phi}\left(\phi\right)}{2B\left(\phi\right)}\right) - V_{,\phi}\left(\phi\right),
\label{eq:eom}
\end{multline}
where we define
\begin{equation}
\gamma^{2}\equiv\frac{B}{A+B\phi^{,\mu}\phi_{,\mu}}.
\end{equation}
See \citet{dis4} for details on the derivation.  Here, a dot represents a partial derivative with respect to cosmic time. The Einstein frame metric is assumed to be a flat Friedmann–Lema\^itre–Robertson–Walker metric with a scalar perturbation $\Psi$, specifically
\begin{equation}
\mathrm{d}s^{2}=-\left(1+2\Psi\right)\mathrm{d}t^{2}+a^{2}\left(t\right)\left(1-2\Psi\right)\left(\mathrm{d}x^{2}+\mathrm{d}y^{2}+\mathrm{d}z^{2}\right).
\end{equation}
The symbol $a$ is the expansion factor, $H=\frac{\dot{a}}{a}$ is the Hubble parameter, and $\rho$ is the total matter density. Note that we have not neglected the non-static terms in the scalar field since, in this particular model, there is a coupling between the density and the time derivatives, whose effects are still not fully understood in the non-linear regime of cosmological evolution.

For convenience, we normalize the field to the vacuum expectation value of the symmetron field,
\begin{equation}
\phi_{0}\equiv\frac{\mu}{\sqrt{\lambda}}.
\end{equation}
As such, the new dimensionless field $\chi=\phi/\phi_{0}$ should stay in the range $\chi\in[-1,1]$, at least for a symmetron-dominated case when $B_{0}$ is small. Also, for numerical convenience, we introduce the parameter $a_{\mathrm{SSB}}$, which defines the expansion factor at the time of spontaneous symmetry breaking, assuming a uniform matter distribution and no disformal term. We further define the density at which the symmetry is broken:
\begin{equation}
\rho_{\mathrm{SSB}}\equiv M^{2}\mu^{2}=\frac{\rho_{0\left(z=0\right)}}{a_{\mathrm{SSB}}^{3}},
\end{equation}
a dimensionless symmetron coupling constant
\begin{equation}
\theta\equiv\frac{\phi_{0}M_{\mathrm{Pl}}}{M^{2}},
\end{equation}
the range of the symmetron field in vacuum
\begin{equation}
\lambda_{0}\equiv\frac{1}{\sqrt{2}\mu},
\end{equation}
and a dimensionless disformal coupling constant
\begin{equation}
b_{0}\equiv B_{0} H_{0}^{2} M_{\mathrm{Pl}}^{2}.
\end{equation}

By taking into account these definitions, we can rewrite Eq. \eqref{eq:eom} as
\begin{multline}
\left(1+\gamma^{2}\rho\right)\ddot{\chi}+3H\dot{\chi}-\frac{1}{a^{2}}\nabla^{2}\chi= \\
\gamma^{2}\rho\left(\frac{4\theta^{2}\zeta}{A\left(\phi\right)}\chi\dot{\chi}^{2}-\frac{\beta}{2}\dot{\chi}^{2}-\frac{1}{2\zeta B\left(\phi\right)M_{\mathrm{Pl}}^{2}}\chi+\frac{1}{a^{2}}\sum_{i=1,2,3}\Psi_{,i}\chi_{,i}\right) \\
+\left(\chi-\chi^{3}\right)\frac{1}{2\lambda_{0}^{2}}, 
\end{multline}
where we have also fixed the three free functions as stated above and defined
\begin{equation}
\zeta\equiv\frac{3\Omega_{0}H_{0}^{2}\lambda_{0}^{2}}{a_{SSB}^{3}}.
\end{equation}

What remains to be done, before we can insert this equation into \noun{Isis}, is to switch to supercomoving time -- the time variable used by \noun{Isis}/\noun{Ramses}, defined by \citet{1998MNRAS.297..467M} -- and to split this second order differential equation into a set of two coupled first-order differential equations. Supercomoving time $\tau$ is related to the cosmic time $t$ by $\mathrm{d}\tau=\frac{1}{a^{2}}\mathrm{d}t$.  Finally, we introduce the variable 
\begin{equation}
q=a\chi', 
\end{equation}
which leads to the following set of first-order differential equations for $q$ and $\chi$:
\begin{align}
\chi' = &\frac{q}{a},\label{eq:unitlessEOM} \\
\begin{split}
q' = &\frac{1}{1+\gamma^{2}\rho} \times
 \left[\rule{0cm}{0.7cm}a^{3}\nabla^{2}\chi \right.\\
& \left. + \gamma^{2}\rho\left(3\tilde{H}q+ \left[\frac{4\theta^{2}\zeta}{A}\chi-\frac{\beta}{2}\right]\frac{q^{2}}{a}+a\sum_{i=1,2,3}\tilde{\Psi}_{,i}\chi_{,i}\right) \right. \\
& \left. + \left(1-\frac{1}{A+B\phi^{,\mu}\phi_{,\mu}}\frac{a_{SSB}^{3}}{a^{3}}\frac{\rho}{\rho_{0}}-\chi^{2}\right)\chi\frac{a^{5}}{2\lambda_{0}^{2}}\rule{0cm}{0.7cm}\right],
\end{split}
\end{align}
where the primes denote derivatives with respect to supercomoving time.
The supercomoving variables with tildes are defined as $\tilde{H}\equiv a^{2}H$ and $\tilde{\Psi}\equiv a^{2}\Psi$.  Setting $\gamma^{2}\rho=B=0$ and $A\approx1$ here will result in symmetron equations equivalent to equations (22) and (23) of \citet{2014PhRvD..89h4023L}.

These equations are solved using the leapfrog algorithm.  The time scales associated with the oscillations of the field are much smaller that those associated with the movement of matter.  Because of this, we use shorter leapfrog time steps for the scalar field than for matter, as described in \citet{2014PhRvD..89h4023L}.

\subsection{The geodesic equation}

In this theory of gravity, dark matter particles move in geodesics determined by the Jordan frame metric. Generally the geodesic equation reads
\begin{equation}
\ddot{x}^{\mu}+\bar{\Gamma}_{\alpha\beta}^{\mu}\dot{x}^{\alpha}\dot{x}^{\beta}=0,
\end{equation}
where $\bar{\Gamma}_{\alpha\beta}^{\mu}$ are the Christoffel symbols that correspond to the Jordan frame metric. Assuming nonrelativistic particles, we neglect quadratic terms in the velocity. The geodesic equation then takes the form
\begin{equation}
\ddot{x}^{i}+\bar{\Gamma}_{00}^{i}+2\bar{\Gamma}_{j0}^{i}\dot{x}^{j}=0.
\end{equation}
To find the necessary barred Christoffel symbols, we use the equation from \citet{2013PhRvD..87h3010Z},
\begin{equation}
\bar{\Gamma}_{\alpha\beta}^{\mu}=\Gamma_{\alpha\beta}^{\mu}+\frac{1}{2}\bar{g}^{\mu\nu}\left[\nabla_{\alpha}\bar{g}_{\beta\nu}+\nabla_{\beta}\bar{g}_{\alpha\nu}-\nabla_{\nu}\bar{g}_{\alpha\beta}\right].
\end{equation}
The resulting equation of motion for the dark matter particles is given by
\begin{align}
\begin{split}
\ddot{x}^{i}+\frac{\Psi_{,i}}{a^{2}}-\frac{2}{AM^{2}a^{2}}\gamma^{2}\phi\phi_{,i}\dot{\phi}^{2}+2\left(H+\frac{\phi\dot{\phi}}{AM^{2}}\right)\dot{x}^{i}\\
+\frac{1}{a^{2}}\gamma^{2}\left(\ddot{\phi}-\frac{1}{a^{2}}\sum_{k=1,2,3}\Psi_{,k}\phi_{,k}+\frac{1}{2}\frac{\beta}{\phi_{0}}\dot{\phi}^{2}\right)\phi_{,i}\\
+2\frac{1}{a^{2}}\gamma^{2}\left(\dot{\phi}_{,j}-H\phi_{,j}-\Psi_{,j}\dot{\phi}+\frac{\beta}{2\phi_{0}}\dot{\phi}\phi_{,j}\right)\phi_{,i}\dot{x}^{j}\\
+\frac{1}{M^{2}a^{2}}\frac{1}{A+B\phi^{,\mu}\phi_{,\mu}}\phi\phi_{,i}-4\gamma^{2}\frac{\phi}{a^{2}AM^{2}}\phi_{,i}\phi_{,j}\dot{\phi}\dot{x}^{j} & = 0.
\end{split}
\end{align}
Here it is possible to recognise the acceleration terms associated with perturbed FLRW geodesics in standard gravity, namely 
\begin{equation}
\ddot{x}^{i}+\frac{\Psi_{,i}}{a^{2}}+2H\dot{x}^{i}=0,
\end{equation}
where the second term is the standard gravity force and the third term is the Hubble friction. These terms are already included in \noun{Ramses}, so we will focus on the other terms.

The effect of modified gravity on the trajectory of the particles is through the addition of a fifth force as well as new damping terms that are proportional to $\dot{\mathbf{x}}$.  As a first attempt to simulate this model, we neglect the damping terms and focus our analysis on the impact of the extra terms associated with the fifth force on a slowly moving matter distribution.  The only friction term that we keep is the one associated with the expansion of the universe, which is automatically taken into account in the solution when writing the equation on supercomoving form. 
The extra acceleration that results from the fifth force, using supercomoving time and with the dimensionless field $\chi$, that we inserted into the code is given by 
\begin{align}
\begin{split}
\label{eq:fifthforce}
x_{\mathrm{fifth}}^{\prime\prime i} & = 
-\frac{2\theta^{2}\zeta}{A+B\phi^{,\mu}\phi_{,\mu}} \, \chi_{,i} \times
\left\{ \rule{0cm}{0.8cm} a^{2}\chi \, + \, \frac{2\zeta}{a^{4}}\frac{b_{0}\exp\left(\beta\chi\right)}{H_{0}^{2}} \times \right.\\
& \left. \left( a\left[q'-3\tilde{H}q-a\sum_{k=1}^3\tilde{\Psi}_{,k}\chi_{,k}\right]+\left[\beta-\frac{4\theta^{2}\chi}{A}\right]\zeta q^{2}\right) \rule{0cm}{0.8cm} \right\} .
\end{split}
\end{align}

\subsection{Tests}

To test that our modifications to the code \noun{Isis} were properly implemented, we compare results that were obtained with our code in the pure conformal limit (i.e. when we set $B_0=0$) to those that were presented by \cite{2014PhRvD..89h4023L} using a pure disformal symmetron non-static code.  We also test the dependence of the final results on the initial conditions that are used for the non-static solver.

\begin{figure}
\resizebox{\hsize}{!}{\includegraphics{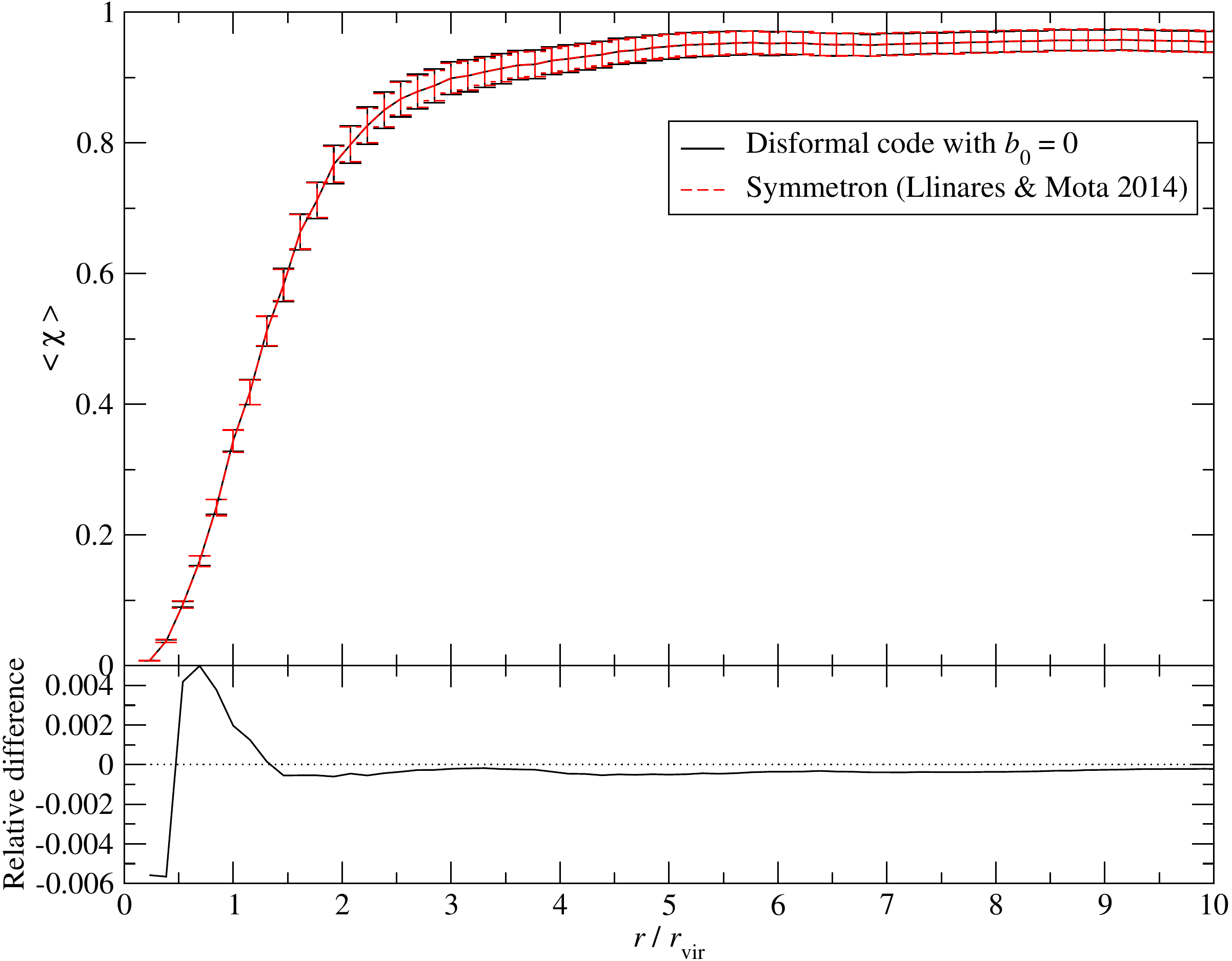}}
\caption{\label{fig:symmcomp-128}Comparison of the average field profiles around a massive halo. The top panel shows the field profile for the symmetron code from \citet{2014PhRvD..89h4023L} (dashed red) and for our disformal code, in the pure conformal limit (solid black). The bottom panel shows the relative difference between the different codes.}
\end{figure}

\begin{figure}
\resizebox{\hsize}{!}{\includegraphics{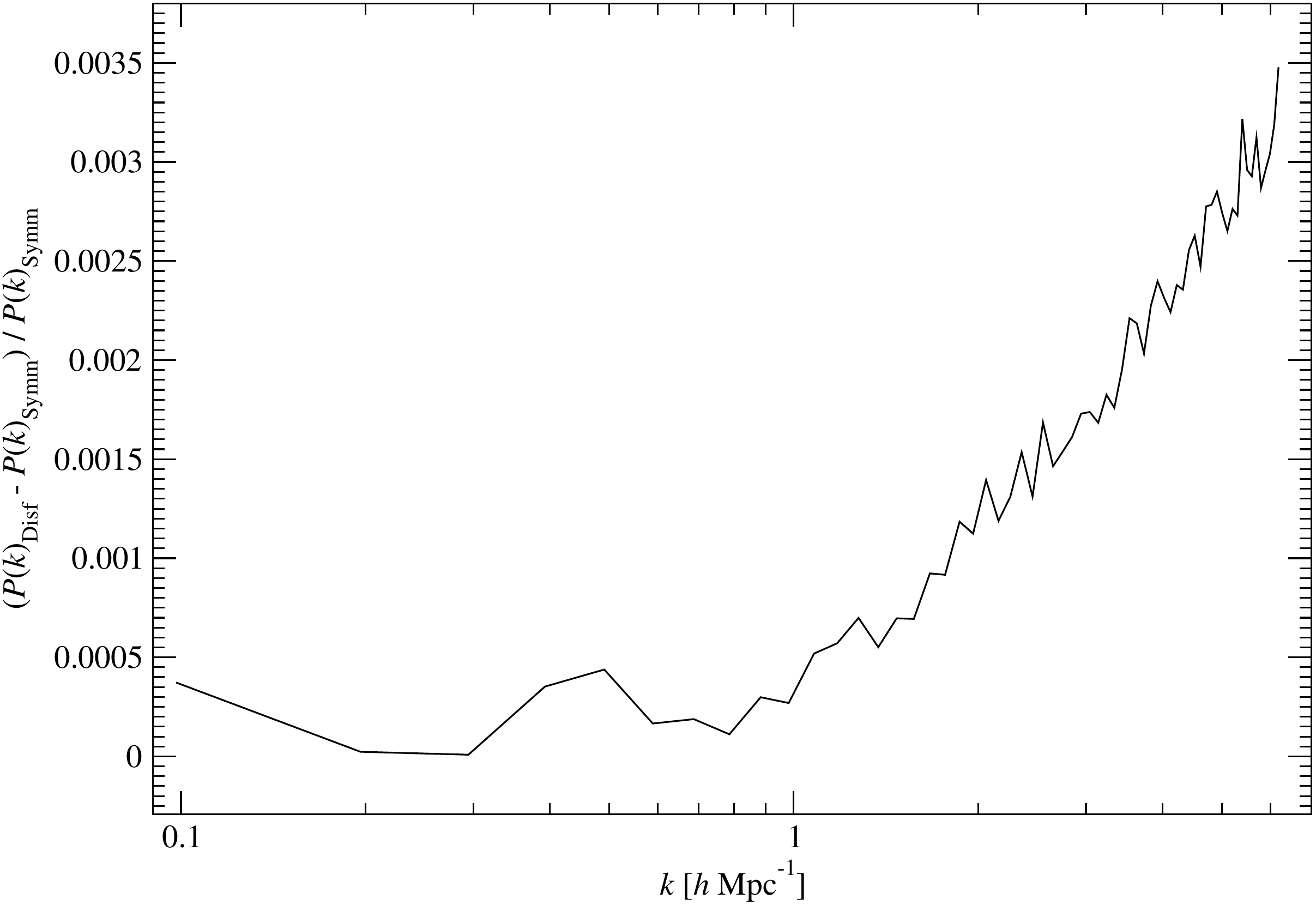}}
\caption{\label{fig:symmcomp-powdiff}Relative difference between the matter power spectra from our disformal code, in the pure conformal limit, and the symmetron code from \citet{2014PhRvD..89h4023L}.}
\end{figure}

\subsubsection{Cosmological comparison to symmetron results}

We run two simulations using both the code presented in this paper and the one described in \citet{2014PhRvD..89h4023L} with the following symmetron parameters: $\theta=1$, $\lambda_{0}=1\,\mathrm{Mpc}$, and $a_{\mathrm{SSB}}=0.5$.  For the disformal part of the model, we used $B_{0}=0$, such that the disformal code should be able to reproduce the already published symmetron non-static results.  Both simulations are done with the same initial conditions, which are generated with the code \noun{Grafic} \citep{2001ApJS..137....1B} using a box size of 64 $\mathrm{Mpc}/h$ and $128^3$ particles.  The comparison of the simulations is focused on the power spectrum of over-density perturbations and the field profiles of the most massive halo.

Figure \ref{fig:symmcomp-128} shows a comparison of the field profile for the most massive halo found in the simulation (with a mass of $2.5\times 10^{14}M_{\odot}/h$).  Because the field oscillates, we compare mean values in time that are calculated by averaging over the time interval from $a=0.995$ to $a=1$.  The differences in the field profiles from the two codes are always below 0.5\% and are sufficiently small to be attributed to the initial random values of the field. A comparison between the power spectra at redshift $z=0$ from both simulations is presented in Fig. \ref{fig:symmcomp-powdiff}.  The relative difference between the power spectra is well below 0.5\% in the whole domain of simulated frequencies. 

\begin{figure}

\resizebox{\hsize}{!}{\includegraphics{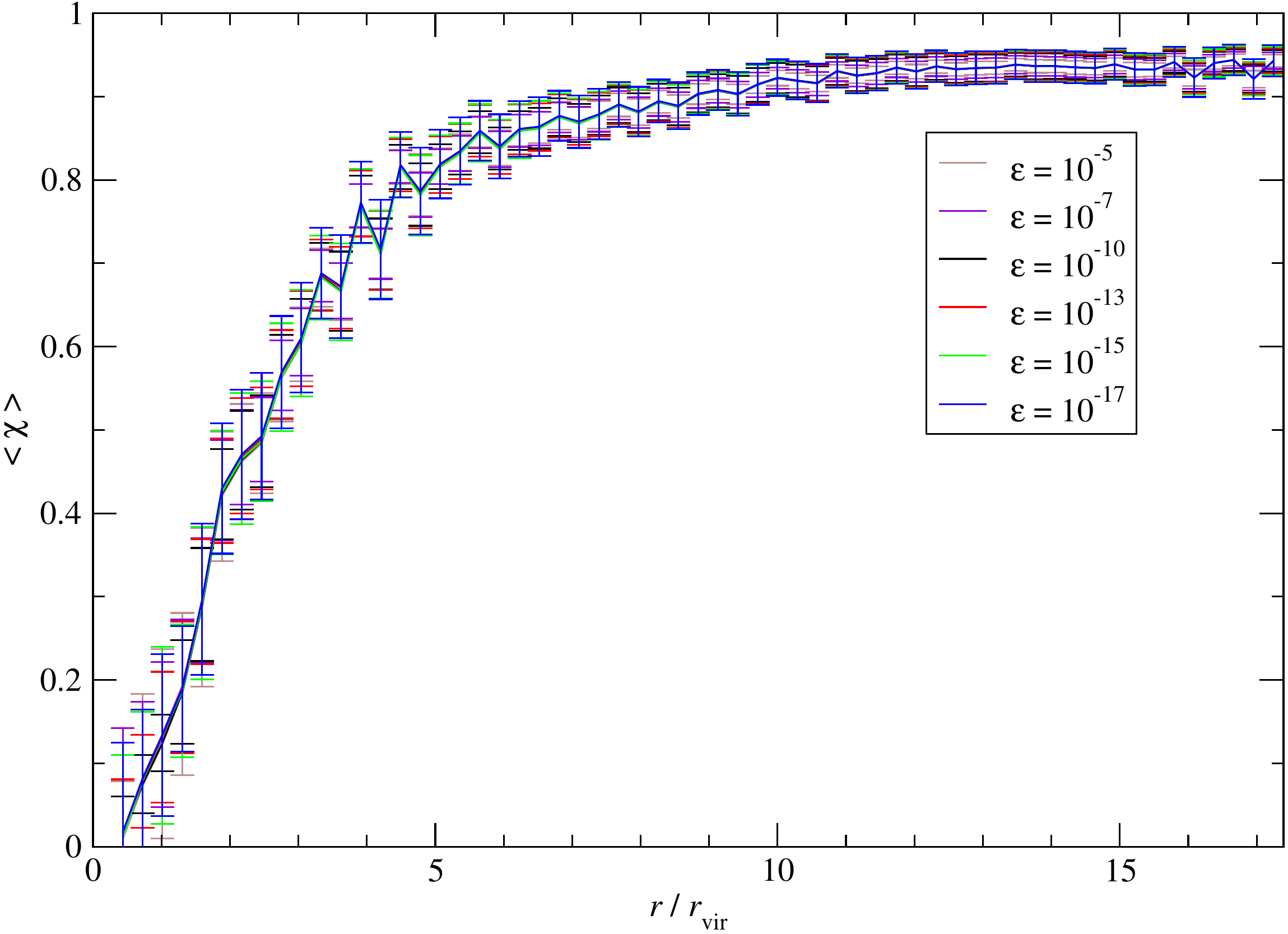}}
\caption{\label{fig:icsig}Sensitivity of the field profiles to differences in the initial conditions for the scalar field.  The curves correspond to the average field profile around a massive halo.  Different curves correspond to different amplitudes $\epsilon$ of the initial random field perturbations.}
\end{figure}

\subsubsection{Dependence on initial conditions for the scalar field}

The initial conditions for the scalar field are generated in the same way as in \cite{2014PhRvD..89h4023L}: the initial values for $\chi$ are drawn from a uniform distribution around zero with a small dispersion ($\chi_{0}\in\left[-\varepsilon,\,\varepsilon\right]$).  We test the sensitivity of the redshift zero scalar field profiles against changes in the amplitude of the initial conditions.  To this end, we run simulations with the only aim of tracking the evolution of the scalar field.  As we are only interested in the field profiles, we use standard gravity in the geodesic equation to ensure identical matter distributions.  The box size and number of particles are 64 $\mathrm{Mpc}/h$  and $64^3$ respectively.  We do several runs changing the amplitude of the initial conditions with values ranging from $10^{-5}$ to $10^{-17}$. Figure \ref{fig:icsig} shows the time averaged scalar field profile that corresponds to the most massive object found at redshift $z=0$ for all these simulations, which has a mass of $1.2\times 10^{14}M_{\odot}/h$.  We find no significant deviations when varying $\varepsilon$.  For the scientific simulations that we present in the following section, we chose to use $\varepsilon=10^{-13}$.

\section{Cosmological simulations}

To quantify the effects of the disformal terms in the cosmological evolution, we run a set of cosmological simulations using Newtonian gravity, a pure symmetron model, and the symmetron model plus the disformal terms.  The initial matter distribution is generated with the package \noun{Grafic} \citep{2001ApJS..137....1B} with standard gravity.  The approximation that we make when not including modified gravity in the initial conditions is fully justified by the fact that because of screening effects, modified gravity starts to act only after the redshift of symmetry breaking has passed, which we choose to be at a much later time.  All simulations use exactly the same initial matter distribution and assume a flat $\Lambda$CDM background cosmology provided by the Planck collaboration: $\Omega_{m}=0.3175$, $\Omega_{\Lambda}=0.6825$, and $H_{0}=67.11$ km/s/Mpc \citep{planck_param}.  The number of particles is $256^{3}$, and the size of the box is 64 $\mathrm{Mpc}/h$.

The initial values of the  dimensionless scalar field $\chi$ at the initial time is calculated by assuming that the scalar field is fully screened at the initial time of the simulation, and thus taken from a uniform random distribution with $\chi_{0}\in[-10^{-13},\,10^{-13}]$. The initial time derivative of the scalar field is assumed to be zero.  As discussed in the previous section, we found that the evolution of the scalar field is not very sensitive to the assumed initial conditions.

\begin{table}
  \centering
  \caption{\label{tab:model_params}Model parameters for the different simulation runs.}

  \begin{tabular}{c c c}
  \hline
  \hline
        Simulation & $b_{0}$ & $\beta$ \\
    
    \hline
    GR (no modifications) & \ldots & \ldots \\
    Symmetron-like & 0 & 0 \\
    Disformal A & 1 & 1 \\
    Disformal B & 2 & 2 \\
    Disformal C & 1 & 0 \\
  \hline
  \end{tabular}
\end{table}

Table \ref{tab:model_params} describes the model parameters employed in the simulations.  The model Disformal A has what we consider to be standard parameters. Disformal B has an amplified disformal part owing to increased $b_{0}$ and $\beta$. Disformal C has $\beta=0$, which sets $B\left(\phi\right)$ to be constant. A constant disformal term might give some insight into the effect of the disformal term alone, since the equations show that the positive and negative parts of the disformal effective potential have different shapes when $\beta\neq 0$. The Symmetron-like simulation is run with the code presented in this paper, but with the disformal part of the equations turned off by setting $b_{0}=0$. In all of the modified gravity simulations, we use the symmetron model parameters $(\lambda_0, a_{SSB}, \theta)=(1 ~\mathrm{Mpc}/h, 0.5, 1)$.

After doing one simulation for each set of parameters, as described in Table \ref{tab:model_params}, we found that Disformal B gave the most unexpected results -- as will be presented in the next section. We therefore performed a total of four simulations of Disformal B, with different initial seeds for the scalar field, but with identical parameters and initial matter distribution. In two of these simulations, the scalar field ended up in the positive minimum of the effective potential ($\phi \approx +\phi_0$), while in two others, the scalar field ended up in the negative minimum ($\phi \approx - \phi_0$).

\section{Results}

\subsection{Power spectrum and halo mass function}

\begin{figure}
\resizebox{\hsize}{!}{\includegraphics{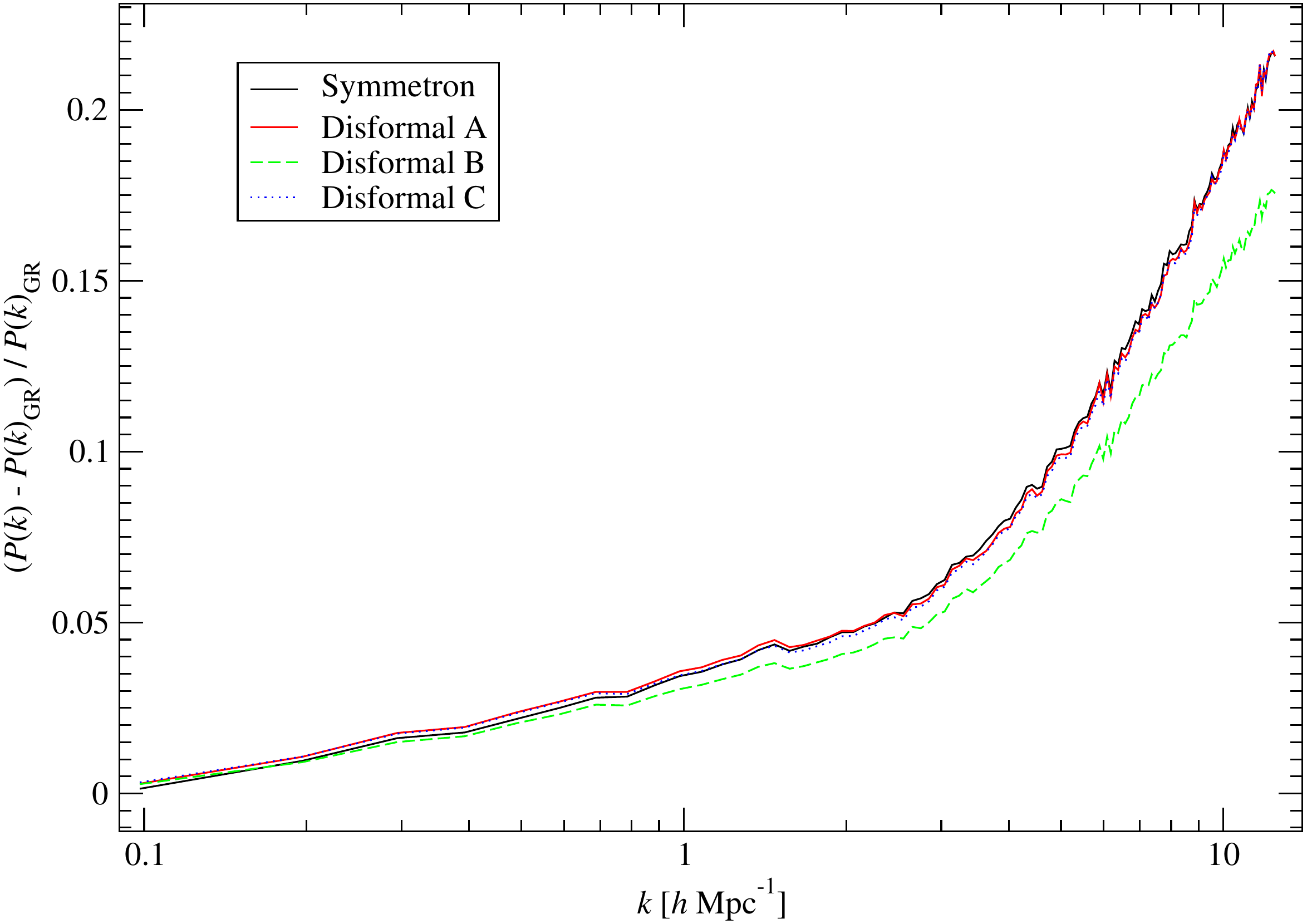}}
\caption{\label{fig:powerspec}Relative difference of the matter power spectrum of the four models of modified gravity with respect to GR.}
\end{figure}

We study the impact of disformal terms in the power spectrum of density perturbations.  The estimation of the power spectrum is made using a Fourier based method.  Discreteness effects are corrected using the method proposed by \citet{2005ApJ...620..559J}.  Figure \ref{fig:powerspec} shows the relative difference between the modified gravity and GR power spectra.  

We find that the pure conformal model has the same effects that were found by \citet{2012ApJ...748...61D} and \citet{2014PhRvD..89h4023L}:  the fifth force has no effect at large scales (which makes the model consistent with the observed normalization of the perturbations) and gives an excess of power at small scales.  Regarding the disformal terms, we found that in the simulations Disformal A and C, the disformal part of the model has no significant impact on the modifications made to GR by the symmetron model.  In the case of the first simulation of Disformal B, the stronger disformal terms counteract the effects of the symmetron, bringing the power spectrum closer to GR.

\begin{figure}
\resizebox{\hsize}{!}{\includegraphics{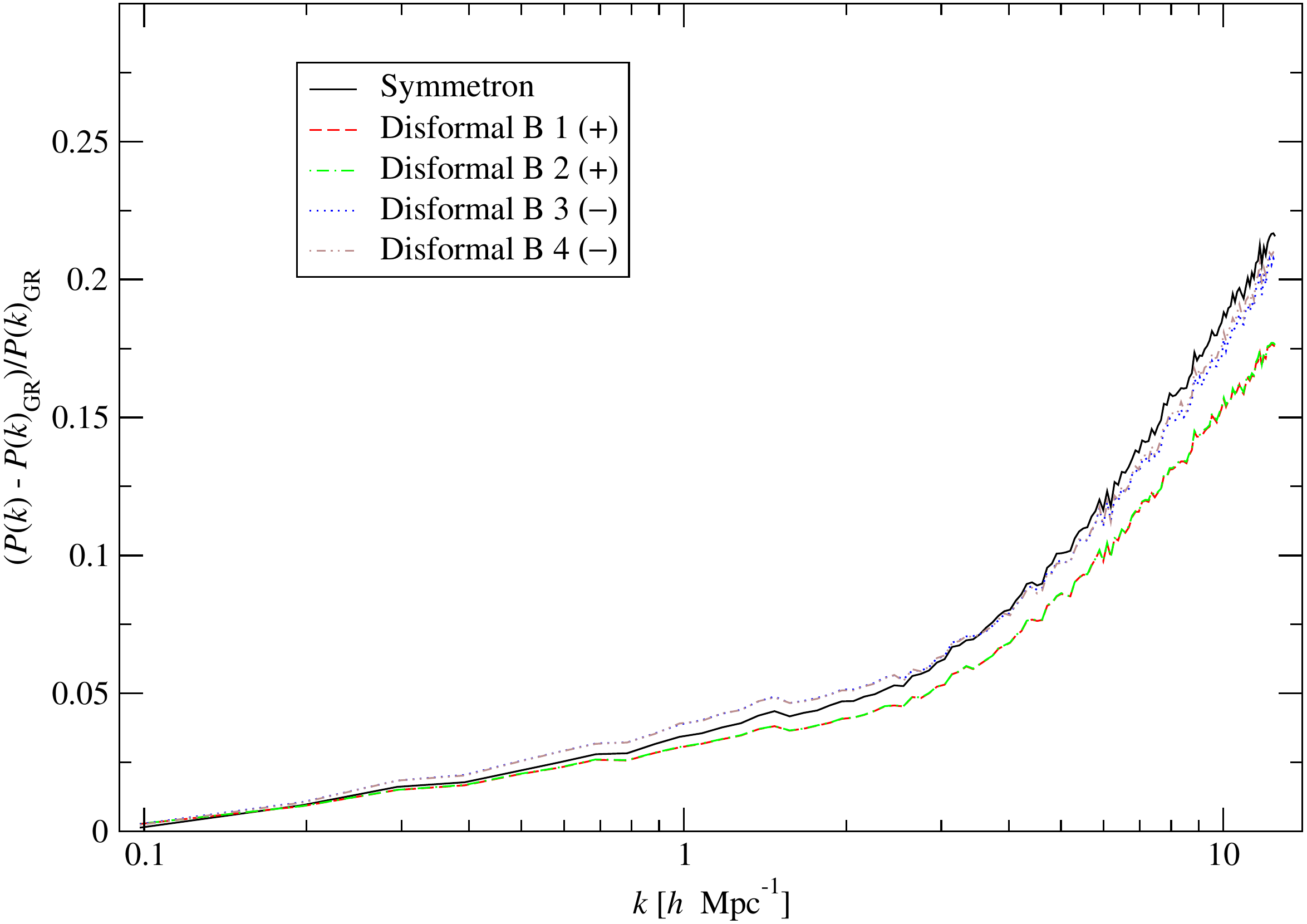}}
\caption{\label{fig:powerspecx}Relative difference of the matter power spectrum of four different simulations of Disformal B with respect to standard gravity (GR). The legend indicates the sign of the potential minimum to which the scalar field fell.}
\end{figure}

Figure \ref{fig:powerspecx} shows the relative difference between all four simulations of Disformal B, and the GR power spectrum. The two disformal simulations where the field fell to the positive potential minimum, have the strongest suppression of power on all scales. For the two other disformal simulations, where the field stabilized around the negative potential minimum, there is some reduction on small scales ($k > 3 h \mathrm{Mpc}^{-1}$), but actually an increase on large scales ($k \approx 1 h \mathrm{Mpc}^{-1}$).

\begin{figure}
\resizebox{\hsize}{!}{\includegraphics{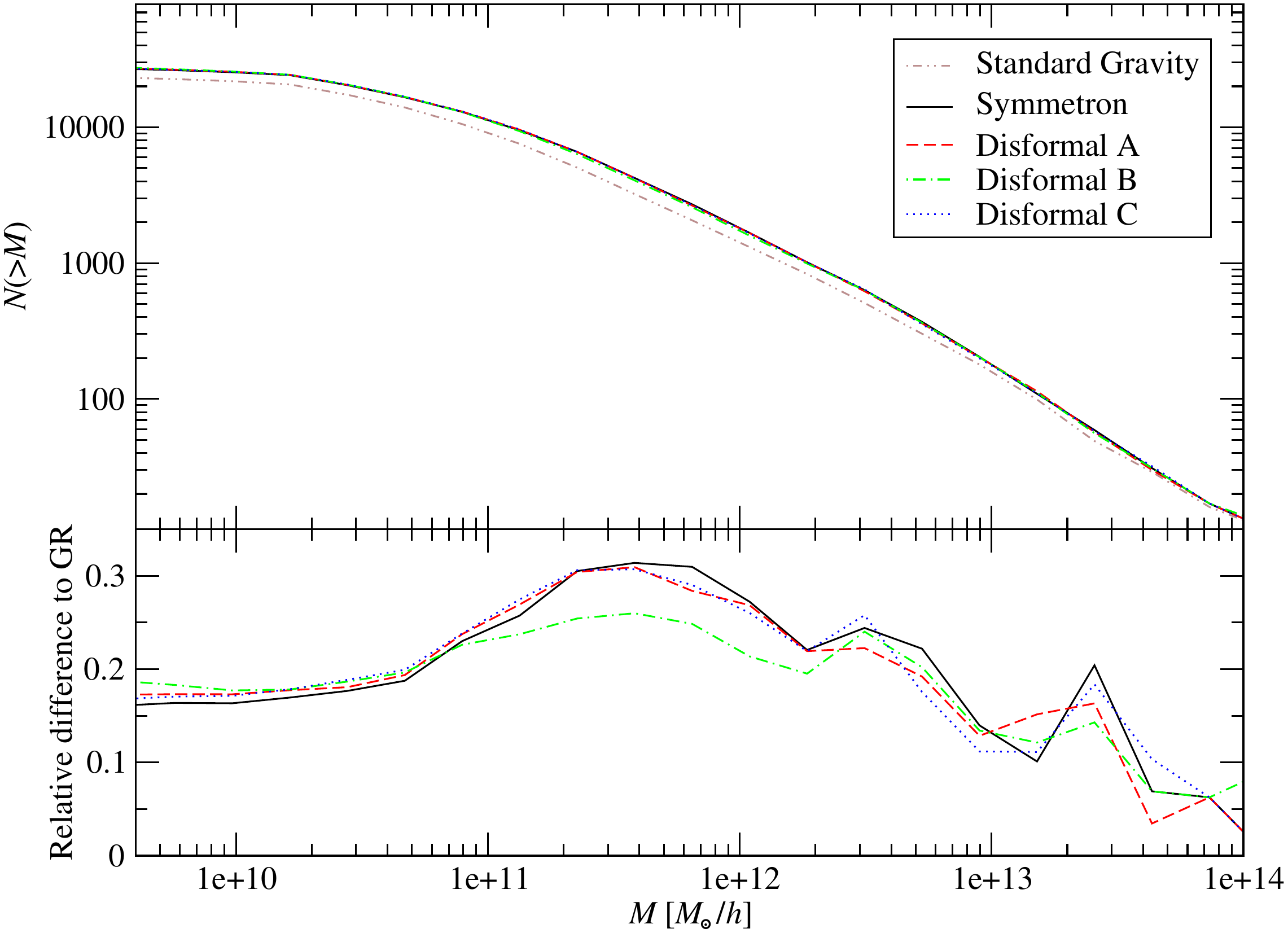}}
\caption{\label{fig:massfunc}Cumulative mass function for each of the five simulations (top panel). The relative difference of the four modified models compared to GR is shown in the bottom panel.}
\end{figure}

Additionally, we study the halo mass function.  To extract this quantity from the simulation data, we identify the dark matter haloes with the halo finder \mbox{\noun{Rockstar}} \citep{2013ApJ...762..109B}, which uses a 6D friends-of-friends algorithm.  The cumulative mass functions for all the simulations are presented in Fig. \ref{fig:massfunc}.  The behaviour is the same as that found in the power spectrum:  the symmetron model increases the number of small haloes, while the strong disformal term in Disformal B acts in the opposite direction, diminishing the symmetron effect.  

These findings on non-linear scales agree with earlier results from linear evolution in disformal theory. In the paper by \citet{2015JCAP...04..036V}, they find on the linear perturbation level, that disformal terms counteract the conformal coupling to some degree. 

\subsection{Field profiles}

We present the radial profile of the scalar field for the most massive halo found in the simulation at redshift $z=0$, which has a mass of $M=5.1\times 10^{14}M_{\odot}/h$.  In the case of static simulations, it is enough to extract information from the last snapshot that is output by the simulation code.  However, in the case of non-static simulations like those presented in this paper, the scalar field has oscillations which makes it difficult to make a comparison between different simulations (the scalar field will often be in a different phase at a given time).  To overcome this problem, we calculate a mean value in time taking into account several oscillation periods.  The mean is calculated over the interval from $a=0.995$ to $a=1$.  We assume here that the variations produced by the displacement of matter during that interval are much smaller than those related to the oscillations of the field.

\begin{figure}
\resizebox{\hsize}{!}{\includegraphics{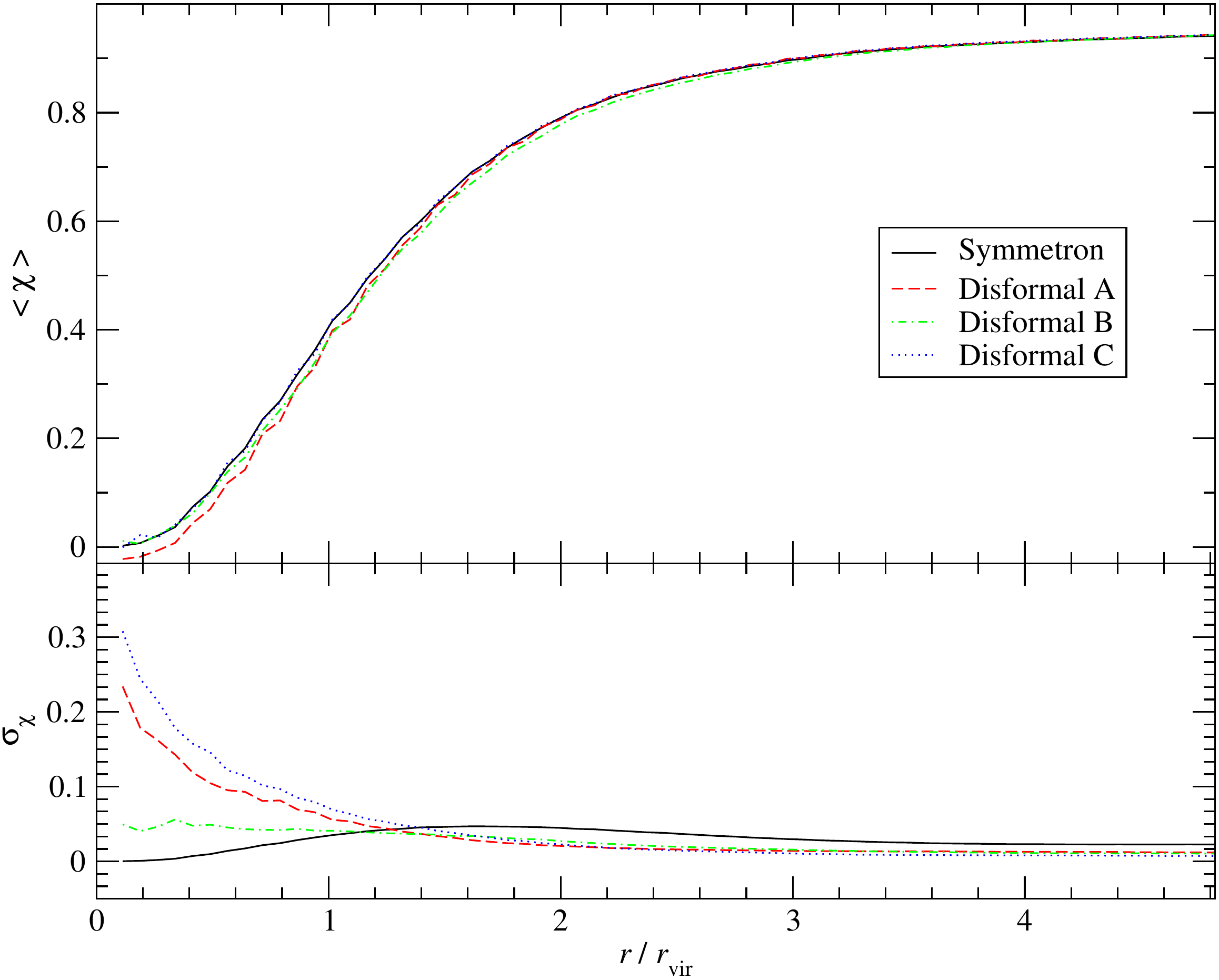}}
\caption{\label{fig:fieldprof}Radial field distribution for the most massive halo found in the simulations. The top panel shows the average field value at a distance $r$ from the centre of the halo. The bottom panel shows the dispersion $\sigma_\chi$ obtained over several oscillations of the scalar field, which is a good estimation of the amplitude of the oscillations.  The field is time-smoothed from $a=0.995$ to $a=1$.}
\end{figure}

The top panel of Fig. \ref{fig:fieldprof} shows the mean profile of the scalar field. In the Symmetron-like, Disformal A, and Disformal C simulations, the scalar field chose to oscillate around the negative minimum (the symmetron potential has two different minima with different signs). In these cases we show $-\langle\chi\rangle$ for a better comparison of the shape of the field profile.
The fact that the potential has two minima should lead to the formation of domain walls \citep{2013PhRvL.110p1101L, 2014PhRvD..90l4041L, 2014PhRvD..90l5011P}, which we do not find in our simulations owing to the small size of the box. Domain walls might have formed in our box at an earlier stage, however, and collapsed before redshift zero.

The scalar field profile tends to the vacuum value far from the halo and presents a gradient responsible for a fifth force when approaching the halo.  The innermost region of the cluster is screened, and thus its corresponding field value tends towards zero in that region.  The profile shows almost no variation among the different models.  However, we find significant differences between standard and disformal symmetron models in the dispersion of the field profiles, which is a measurement of how large the amplitudes of the scalar field oscillations are.  
This information is shown in the bottom panel of Fig. \ref{fig:fieldprof}.  In the symmetron case, we find that the dispersion of the profile goes to zero in the centre of the cluster, which means that the scalar field is completely at rest there and only has oscillations in the outer regions and the voids surrounding the halo.  In the disformal case, the situation is different. We find that the amplitude of the oscillations does not go to zero in the centre of the cluster and that, in fact, the amplitude can even increase in the central region with respect to the values found outside of the halo.  We also see that the dispersion is slightly less than in the symmetron case in the low density region ($r > 1.5 r_{\mathrm{vir}}$).  The reason for this may be related to the speed of waves of the scalar field being less in high density regions (see Eq. \eqref{eq:eom}, which has the form of a wave equation where the factor of $(1 + \gamma^2 \rho)$ in front of $\ddot{\phi}$ can be regarded as $1/c_s^2$). Thus, inside massive haloes, the waves of the scalar field can cluster and give rise to extra oscillations.  Further analysis must be made to confirm this hypothesis.

\section{Conclusions}

For the first time, we run cosmological simulations with disformally coupled scalar fields.  The model includes both a conformal and a disformal coupling to matter.  For the conformal part we choose a symmetron potential and a conformal factor $A = 1+\left(\phi / M \right)^{2}$.  The disformal part is given by an exponential factor $B = B_{0}\exp\left(\beta \phi / \phi_{0}\right)$.  The aim of the paper is to test the effects of the disformal factor on the already known results for the symmetron model.

We present the formalism and modifications that we made to the code \noun{Isis} \citep{2014A&A...562A..78L} to be able to simulate disformal gravity.  The disformal code presented here is based on the non-static solver for scalar fields that is part of the \noun{Isis} code \citep{2014PhRvD..89h4023L}.  We present results obtained from a set of five cosmological simulations.

In the set of simulations, we find that the stronger disformal terms can counteract some of the clustering effects of the symmetron field: the power spectrum of density perturbations and the halo mass functions are both smaller than in the symmetron case.  However, at least in the region of the parameter space studied, the conformal part of the model is dominant. Therefore, the end result is nevertheless a small-scale increase of both the power spectrum and the mass function with respect to GR.

In the first simulation we found that Disformal B was the only model where the field fell to the positive minimum of the symmetron potential. This could be very important, since the disformal factor $B \left( \chi \right) \propto b_0 \exp \left( \beta \chi \right)$ will change by several orders of magnitude when $\chi \rightarrow -\chi$.  The reason for the field falling to one minimum or the other in a particular simulation is complex and can be attributed to the chaotic nature of the equation of motion for the field. We did a total of four simulations on Disformal B with varying initial field distributions. The results show that the power spectrum was reduced significantly only for positive field values.  This indicates that the exponential shape of $B(\phi)$ is more important than the value of $b_0$ to achieve the masking of the symmetron clustering. Furthermore, we will probably see new physics and observational signatures of this model in future studies of domain walls \citep[see][for a description of domain walls in asymmetric potentials]{2014PhRvD..90b3521C}.

To understand the differences in the distribution of 
dark matter that appear because of the new terms, we study the field profiles that correspond to the most massive halo found in the simulations.  We find almost no differences in the field profile, but we do find differences in the amplitude of the oscillations of the field, which are larger for the disformaly coupled models. This implies that the differences in the power spectra and mass functions when comparing the disformal models and the symmetron are not the result of the field value, but of the field oscillations.  While the gradient of the scalar field is independent of the presence of the disformal terms, it is important to keep in mind that the expression for the fifth force includes the time derivatives of the scalar field.  Hence, the fifth force can be modified with respect to the purely conformal fifth force by only increasing the derivatives.

Possible observational consequences of the field oscillations in haloes are (time varying) changes in how photons are lensed by the oscillating disformally coupled field. Because light rays will follow geodesics that are dictated by the modified Jordan frame metric, anomalies in cluster lensing masses could indeed be a `disformal smoking gun'. This possibility will be studied in a future paper.

\begin{acknowledgements}

The simulations were performed on the NOTUR cluster HEXAGON, the computing facilities at the University of Bergen.  CLL and DFM acknowledge support from the Research Council of Norway through grant 216756.
\end{acknowledgements}

\bibliography{references}

\end{document}